\documentclass[aps,preprint,showpacs]{revtex4}

\usepackage{graphicx}
\usepackage{bm}

\begin{document}

\title{Microscopic spectral density in random matrix models for chiral
and diquark condensation}

\author{Beno\^\i t Vanderheyden} \affiliation{Departement
of Electrical Engineering and Computer Science, \\
B-28, Universit{\'e} de Li\`ege, Sart-Tilman, \\ 
B-4000 Li\`ege, Belgium}

\author{A. D. Jackson}
\affiliation{
The Niels Bohr Institute, Blegdamsvej 17, DK-2100 Copenhagen \O, Denmark}

\date{\today}

\begin{abstract}
  
  We examine random matrix models of QCD which are capable of
  supporting both chiral and diquark condensation. A numerical study
  of the spectral densities near zero virtuality shows that the
  introduction of color in the interactions does not alter the
  one-body results imposed by chiral symmetry. A model with three
  colors has the spectral density predicted for the chiral ensemble
  with a Dyson index $\beta = 2$; a pseudoreal model with two colors
  exhibits the spectral density of the chiral ensemble with $\beta =
  1$.

\end{abstract}

\pacs{11.30. Fs, 11.30. Qc, 11.30. Rd, 12.38. Aw}

\maketitle

Chiral random matrix theory ($\chi$RMT) is based on the observation
that many of the low-energy properties of QCD are dominated by its
global symmetries~\cite{reviewchiRMM}. Random matrix
models~\cite{chiRMM} thus attempt to capture the basic mechanisms for
chiral condensation by reducing the QCD interactions to their
essential structure. These models introduce low-lying modes which
respect a basic left-right symmetry but which interact via random
matrix elements.  The consequences of chiral symmetry have been
investigated at two levels. The microscopic level deals with the
statistical properties of the eigenvalues of the Dirac operator and
their correlations.  Here, $\chi$RMT
studies~\cite{VerZah93,Ver94,Ver94-2} have helped in understanding how
these properties are determined by the spontaneous breaking of chiral
symmetry as indicated, for example, by the universality of the
spectral density near zero virtuality and associated sum
rules~\cite{reviewchiRMM}.  At the macroscopic level, it is possible
to consider the consequences of chiral symmetry for the global state
of the random matrix system.  With the aid of the mean field
approximation and additional prescriptions for including the effects
of temperature and chemical potential, it is straightforward to
construct the partition function for $\chi$RMT models and thus examine
the patterns of chiral symmetry breaking as a function of $T$, $\mu$,
and the quark mass~\cite{JacVer96,HalJac98}.  Many of the properties
of the resulting phase diagram are direct consequences of chiral
symmetry and largely independent of the detailed form of the
interactions.  They are thus expected to provide guidance in our
understanding of the QCD phase diagram particularly in cases (e.g.,
$N_c = 3$ and $\mu \ne 0$) where the Dirac operator is non-Hermitean.
These are cases where lattice simulations cannot rely on importance
sampling and are very difficult to
perform~\cite{AlfKap99,ChaWie99,FodKat02,ForPhi02}.

There has been considerable interest in the possibility that both
chiral and diquark condensates can develop and compete
thermodynamically in QCD. (For a review of diquark condensation in
QCD, see~\cite{early,RapSch98,AlfRaj98,review}.)  Thus, we recently
proposed a random matrix model for QCD which goes beyond $\chi$RMT and
has interactions which implement coexisting chiral and color
symmetries~\cite{VanJac99,VanJac01}.  We have examined this model at
the macroscopic level by identifying the allowed topologies for the
$(T,\mu)$ phase diagram.  The phase structure is determined by a
single parameter $\alpha$, defined as a ratio between coupling
constants in the chiral and diquark condensation channels. Physically,
this ratio measures the relative strengths of the chiral and color
symmetries in the random matrix interactions.  In the cases relevant
for QCD [either ${\rm SU}(2)$ or ${\rm SU}(3)$], $\alpha$ has a fixed
value and is associated with a given phase structure. The topology of
this phase structure is robust with respect to moderate variations in
the detailed form of the interactions (i.e., variations in $\alpha$).

The addition of color structure to the interactions implies additional
constraints on the random matrix ensemble considered.  As we shall see
below, the color generators modify the statistical weights of the
interaction matrix elements. The question then arises whether these
constraints are capable of altering the various results associated
with chiral symmetry alone.  Our previous considerations suggest that
this is not the case at the macroscopic level.  In the case of an
interaction which is completely dominated by chiral symmetry ($\alpha
\to 0$), our models precisely reproduce the phase diagram of $\chi$RMT
(Ref.~\cite{JacVer96}). In cases relevant for QCD with fundamental
fermions~\footnote{In this work, we consider only fundamental
fermions. For models with adjoint fermions,
see~\cite{reviewchiRMM,KogSteTou00,KimSon01,KimVer01}} and three
colors ($N_c = 3$ and $\alpha = 0.75$), the diquark phase develops in
regions where it is thermodynamically advantageous, but color does not
otherwise modify the phase structure.  Thus, for $N_c = 3$, color does
not seem to weaken the chiral correlations at the macroscopic level.

For $N_c = 2$, the situation is more subtle: the gauge interaction is
pseudoreal, and baryons and mesons belong to the same multiplets. For
$m =0$, $\mu = 0$, and $N_f$ flavors, the quark Lagrangian has an
extended ${\rm SU}(2 N_f)$ invariance which relates chiral and diquark
fields. This flavor symmetry is explicitly broken for either $m >0$ or
$\mu >0$. Studies of the symmetry breaking patterns showed that a
diquark condensed phase becomes favorable for $\mu \sim m_\pi \propto
m^{1/2}$ and $T <
T_c$~\cite{KogSteTou99,KogSteTou00,KimSon01,KimVer01}. Our model
exhibits the ${\rm SU}(2 N_f)$ symmetry and produces results which agree
with the general predictions of Ref.~\cite{KogSteTou99} as well as
with those of chiral perturbation
theory~\cite{KogSteTou00,KimVer01}. Thus, for $N_c = 2$, the
introduction of color in the random matrix interactions does not lead
to unexpected topologies of the phase diagram (i.e. unexpected
macroscopic properties).

The purpose of this paper is to investigate the consequences of the
additional color correlations at the microscopic level. To this end,
we examine the distributions of the eigenvalues of the Dirac operator
near zero virtuality and compare them with the analytic forms
predicted by $\chi$RMT. The rest of the paper is organized as follows:
We present the models in Sec.~\ref{s:models}, discuss aspects of their
macroscopic and microscopic spectral densities in
Secs.~\ref{s:macrosd} and \ref{s:microsd}, comment on other properties
in Sec.~\ref{s:others}, and conclude in Sec.~\ref{s:concl}.

\section{The random matrix models}
\label{s:models}

To understand the form of the correlations induced by each symmetry,
it is useful to provide a brief description of the models of
Refs.~\cite{VanJac99}, which we will refer to as I. In the present
paper we will restrict our attention to a theory with $N_f$ light
flavors, zero temperature, zero chemical potential, and zero quark
mass.  We start by recalling the basic form of chiral random matrix
theory, in which only chirality is introduced, and then extend the
model to include color.

We consider first a chiral random matrix model and work in the sector
of zero topological charge for simplicity.  The partition function has
the form
\begin{eqnarray}
Z = \int\,{D} W \,\,\prod_{i = 1}^{N_f}\,\, {D}\psi_i^*\,{D}
\psi_i^{\phantom{*}} \, 
\exp\left[i \sum_{i= 1}^{N_f} \, \psi^*_i \, {\cal D} \, \psi_i
\right]\,\exp\Big(- \frac{N \beta \Sigma^2}{2}\,{\rm Tr}[W W^\dagger]\Big),
\label{Z}
\end{eqnarray}
where $\psi_i^{\phantom *}$ and $\psi^*_i$ are independent Grassmann
variables representing the quark fields and where the matrix ${\cal
D}$ represents the Dirac operator. Its block structure reflects chiral
symmetry. Working in a suitable basis of left and right states $(1 \pm
\gamma_5) \phi_n$, ${\cal D}$ has the form~\cite{chiRMM}
\begin{eqnarray}
{\cal D} & = & \left( \begin{array}{cc} 0 & W \\ W^\dagger & 0
        \end{array} \right),
\label{D}
\end{eqnarray}
where $W$ is an $N\times N$ block matrix. The integral in
Eq.~(\ref{Z}) is over the matrix elements of $W$, $DW$ is a Haar
measure, and Tr denotes a trace over the $N$ matrix indices. The model
is thus a theory of $2 N$ low-lying modes which respect chiral
symmetry and whose interaction matrix elements $W_{ij}$ are drawn on a
Gaussian distribution. The number of modes scales with the volume of
the system; the thermodynamic limit is taken as $N \to \infty$.

For later comparisons, it is useful to understand how the gauge group
is taken into account. In a model which implements the global
symmetries of QCD with $N_c = 3$, the matrix elements of $W$ are
complex. For the analysis below, it is worth noting that their real
and imaginary parts satisfy no particular relationship and are thus
drawn independently. This case corresponds to a Dyson index $\beta =
2$ and is described by the chiral unitary ensemble ($\chi$GUE). For
$N_c = 2$, the gauge group is pseudoreal as mentioned above.  The
Dirac operator then contains an additional antiunitary
symmetry~\cite{chiRMM}, which allows one to choose a basis in which
the matrix elements of $W$ are real. This case leads to the chiral
orthogonal ensemble ($\chi$GOE).  In short, the value of $N_c$ enters
$\chi$RMT only through the reality of the matrix elements of $W$.

In model I, we included color directly in the interactions in a way
which mimics single-gluon exchange.  For two flavors, the partition
function takes the form
\begin{eqnarray}
Z = \int\,{D} H \,\,{D}\psi_1^\dagger\,{D}\psi_1^{\phantom{*}} \,
 {D}\psi_2^*\,{D}\psi_2^T \,
\exp\left[i 
\left(
\begin{array}{c}
\psi_1^\dagger \\
\psi_2^T \\
\end{array}
\right)^T
\left(
\begin{array}{cc}
{\cal D}_c & 0 \\
0 & -{\cal D}_c^T \\
\end{array}
\right)
\left(
\begin{array}{c}
\psi_1 \\
\psi_2^*
\end{array}
\right)
\right],
\label{Zcolor}
\end{eqnarray}
where $\psi_1$ and $\psi_2^T$ denote the quark fields for flavor $1$
and flavor $2$, respectively, $D H$ is a measure to be defined below,
and ${\cal D}_c$ is the single-quark Dirac operator.  Note that the
subblock associated with flavor $2$ has been transposed in order to
exhibit the possibility of forming $\langle \psi_2^T \psi_1^{\phantom
T} \rangle$ condensates (see I).

The diagonal block ${\cal D}_c$ now reflects both chiral and color
symmetries. In order to be able to define an order parameter which is
antisymmetric under the permutation of an odd number of quantum
numbers, spin has to be introduced together with
color~\cite{VanJac99}.  The Dirac operator ${\cal D}_c$ then has the
chiral structure of Eq.~(\ref{D}) where $W$ is exploded into a $2 N_c
\times 2 N_c$ matrix of embedded spin and color subblocks,
\begin{eqnarray}
W = \sum_{\mu = 0}^3 \, \sum_{a = 1}^{N_c^2 - 1}
(\sigma_\mu \otimes \lambda_a)\, A_{\mu a}. 
\label{W}
\end{eqnarray} 
Here $\sigma_\mu = (1, i \vec{\sigma})$ are $2 \times 2$ spin
matrices, $\lambda_a$ are the $N_c \times N_c$ matrices of ${\rm
SU}(N_c)$ and $A_{\mu a}$ are $n \times n$ real matrices representing
the gluon fields.  Taking into account all substructures, each matrix
$W$ is thus $N \times N$ with $N = 2 N_c n$.  The matrix elements of
$A_{\mu a}$ in Eq.~(\ref{W}) are distributed according to the measure
${D}H$, Eq.~(\ref{Zcolor}), which takes the form
\begin{eqnarray}
{D}H = \left\{ \prod_{\mu a}\,{D}A_{\mu a} \right\}
\,\exp\left(- N \Sigma_0^2 \sum_{\mu a} {\rm Tr}[A_{\mu a} \, 
(A_{\mu a})^T]\right),
\label{DH}
\end{eqnarray}
where $D A_{\mu a}$ are Haar measures, $\Sigma_0$ is a constant, and the
superscript $T$ denotes a transposition.  Again, the thermodynamic limit
corresponds to $N \to \infty$.

Evidently, these spin and color substructures impose strong
constraints between the real and imaginary parts of the matrix
elements of $W$.  Consider, for example, the contributions to ${\cal
D}_c$, Eqs.~(\ref{Zcolor}) and (\ref{W}), of the random matrices
$A_{01}$ ($\mu =0$ and $a = 1$) and $A_{02}$. Their matrix elements
are real and are drawn independently.  When combined to form $W$ as
prescribed in Eq.~(\ref{W}), the matrix elements of $A_{01}$ multiply
$\sigma_0$, a real diagonal spin matrix, and $\lambda_1$, a real color
matrix. Hence, $A_{01}$ contributes to the real part of $W$
only. Similarly, the matrix elements of $A_{02}$ combine with
$\sigma_0$ (real) and $\lambda_2$, an imaginary color matrix. They
thus contribute to the imaginary part of $W$. Hence, in contrast to
ordinary $\chi$RMT, the matrix elements of $W$ are complex for both
$N_c =3$ and $N_c = 2$. The real and imaginary parts of $W$ arise from
well-defined combinations of the matrix elements of $A_{\mu a}$. Their
statistical distributions are then dictated by the content of the spin
and color block matrices, which thus introduce well-defined
correlations.

Do these additional correlations preserve those imposed in
Eq.~(\ref{D}) by chiral symmetry? The $W$ matrix elements are complex
for all $N_c$; their real and imaginary parts are no longer
independent random variables.  One might thus be concerned that the
statistical properties of the eigenvalues of the Dirac operator would
differ from those of the $\chi$GUE and $\chi$GOE.  We now consider a
number of spectral properties to indicate that this is {\em not\/} the
case and that the additional color symmetries do not alter the
statistical properties due to chiral symmetry.

\section{The macroscopic spectral density}
\label{s:macrosd}

As an initial measure of the statistical properties, we consider the
spectral density $\rho(\lambda)$, defined as
\begin{eqnarray}
\rho(\lambda) & \equiv & {1 \over 2 N} \, \left\langle 
\sum_{i = 1}^{2 N} \delta(\lambda - \lambda_i)\right\rangle,
\end{eqnarray}
where $\lambda_i$ are the $2 N$ eigenvalues of the Dirac operator and
$\langle \rangle$ denotes an ensemble average.  While $\rho(\lambda)$ 
is not universal in the usual random matrix sense, we will focus on 
those symmetry properties which are expected to be protected.  
Consider first the chiral random matrix model defined in Eq.~(\ref{Z}).  
Because of the block structure of ${\cal D}$, Eq.~(\ref{D}), the eigenvalues
$\lambda_i$ occur in pairs of opposite signs. Moreover, Eq.~(\ref{Z})
shows that the spectrum is $N_f$-fold degenerate. Consider next model I, 
Eq.~(\ref{Zcolor}), which may seem different at first glance 
because of its more elaborate structure. In fact, each flavor subblock has 
the same chiral substructure as in Eq.~(\ref{D}), and the eigenvalue 
spectrum remains symmetric about $\lambda = 0$.  Further, the eigenvalues 
of $- {\cal D}_c^T$, Eq. (\ref{Zcolor}), are degenerate with those of 
${\cal D}_c$ so that the spectrum is again twofold degenerate ($N_f = 2$).  
Hence, the same basic chiral and flavor symmetries prevail in the two models.

In order to facilitate numerical evaluation, we now consider the
quenched limit $N_f \to 0$. This limit is free from contributions from
vacuum graphs (through powers of the determinant of the Dirac operator
in the partition function; see~\cite{reviewchiRMM}). The ensemble
average of a given quantity then amounts to a mere counting of the
contributions from the individual eigenvalues of ${\cal D}$ (or ${\cal
D}_c$), distributed according to the normal laws of Eq.~(\ref{Z}) or
Eq.~(\ref{DH}), as appropriate.

For large matrices, we find that the model I is numerically consistent
with the (non universal) semicircle law familiar from the $\chi$GUE
and $\chi$GOE,
\begin{eqnarray}
\lim_{N \to\infty} \rho(\lambda) & = & \left\{ \begin{array}{cc}
        \frac{\Sigma}{\pi}\, \sqrt{1 
- \left(\frac{\Sigma \lambda}{2}\right)^2} & 
        \textrm{if}~|\lambda| \le 2/\Sigma, \\
        0 & \textrm{otherwise}. 
                            \end{array}
                    \right.  
\label{rho}
\end{eqnarray}
In the case of chiral random matrix models, $\Sigma$ is the variance
of the distribution in Eq.~(\ref{Z}). In model I, $\Sigma$ is
proportional to $\Sigma_0$, Eq.~(\ref{DH}), in a manner that will be
discussed shortly.  Note that thanks to the Banks-Casher
relationship~\cite{BanCas80}, $\Sigma$ is in all cases to be
identified with the chiral order parameter:
\begin{eqnarray}
\langle \bar\psi \psi\rangle = \lim_{\lambda \to 0}\,\, \lim_{N \to \infty}
\pi \rho(\lambda) = \Sigma.
\end{eqnarray}
Having obtained the semicircle distribution in model I is a natural result.
The random matrix interactions of Eq.~(\ref{W}) mix a set of $4 \times (N_c^2
- 1)$ independent real matrices, $A_{\mu a}$, in a democratic way. This
ensemble naturally leads to the semicircle law familiar from most elementary
random ensembles, including the chiral ones.

Further remarks can be made about the dependence of $\Sigma$ with
respect to $N_c$. For $\chi$RMT, $\Sigma$ does not depend on $\beta$,
and hence does not depend on $N_c$.  In model I, however, $\Sigma$ is
a function of $N_c$ and $\Sigma_0$ which can be easily determined by
noting the following relationship~\footnote{This can be established by
considering the resolvent operator $G(z) = \langle {\rm Tr} (z -
D)^{-1}\rangle$, and by matching its asymptotic expansion at large $z$
to the form it assumes for the spectral density of Eq.~(\ref{rho}).}
between the radius of the semicircle $2/\Sigma$ and the variance
$\langle {\rm Tr}[W W^\dagger]\rangle$:
\begin{eqnarray}
\langle {\rm Tr}[W W^\dagger]\rangle =
\frac{N}{\Sigma^2}.
\label{varChi}
\end{eqnarray}
From the definition of $W$, Eq.~(\ref{W}), and the distribution in
Eq.~(\ref{DH}), we have
\begin{eqnarray}
\langle {\rm Tr}[W W^\dagger]\rangle = 4 \sum_{\mu a} 
\langle {\rm Tr} [A_{\mu a} (A_{\mu a})^T] \rangle =
\frac{2 (N_c^2 - 1)}{N_c^2}\,\frac{N}{\Sigma_0^2},
\label{SigmaSquared}
\end{eqnarray}
which gives
\begin{eqnarray}
\Sigma & = & \frac{N_c}{\sqrt{2 \,(N_c^2 - 1)}} \,\,\Sigma_0 =
\left\{\begin{array}{cc}
        0.75\,\Sigma_0 & \textrm{if}~N_c = 3, \\
        \\
        0.8165\,\Sigma_0 & \textrm{if}~N_c = 2.
       \end{array}
\right.
\label{SigmaColor}
\end{eqnarray}
The $N_c$ dependence of the chiral condensate reflects the fact that
the strength of the interactions in the chiral channel varies with
$N_c$. In fact, the $N_c$-dependent prefactor in
Eq.~(\ref{SigmaSquared}) is directly proportional to the Fierz
coefficient which appears when the random matrix interactions are
projected onto the chiral condensation channel (Ref.~\cite{VanJac99}).

\section{The microscopic spectral density}
\label{s:microsd}

The structures that are characteristic of chiral symmetry are most
clearly seen in the distributions of the eigenvalues near zero
virtuality. It is thus worth considering the microscopic spectral
density, defined at the level of single eigenvalues as
\begin{eqnarray}
\rho_S(z) = \lim_{N \to \infty} {1 \over 2 N \Sigma}\,
\rho\left({z \over 2 N \Sigma}\right),
\end{eqnarray}
where $\Sigma$ is the chiral order parameter of the theory at hand.  The 
microscopic spectral densities of the various chiral ensembles are strikingly 
different.  In the quenched limit, $\chi$GUE gives
\begin{eqnarray}
\rho_S(z) & = & {z \over 2}\,\left(J_0^2(z) + J_1^2(z)\right)
~~~~(\textrm{$\chi$GUE}, N_f = 0, z >0),
\label{rhocGUE}
\end{eqnarray}
while $\chi$GOE gives~\cite{reviewchiRMM,ForNag99,KleVer00} 
\begin{eqnarray}
\rho_S(z) & = & {z \over 2}\, \left( J_1^2(z) - J_0(z) \,J_2(z) \right)
- {1 \over 2} \, J_0(z) \, \left(\int_0^z du \,\,J_2(u) - 1 \right)
\nonumber \\
&& (\textrm{$\chi$GOE}, N_f = 0, z >0).
\label{rhocGOE}
\end{eqnarray}

Figures~1 and 2 show the histograms obtained in the models of
Refs.~\cite{VanJac99}, plotted against Eq.~(\ref{rhocGUE}) for the
model with $N_c = 3$ and against Eq.~(\ref{rhocGOE}) for that with
$N_c = 2$. The data were obtained from an ensemble of $2 \times 10^5$
matrices with blocks $W$ of size $120 \times 120$ (i.e., $2 n N_c =
120$).  As in $\chi$RMT, the behavior of the small eigenvalues depends
strongly on whether $N_c = 3$ or $N_c = 2$.  In fact, the data agree
with the theoretical curves for the corresponding chiral ensembles to
expected statistical accuracy.  The microscopic spectral density thus
appears to preserve the form dictated by chiral symmetry in spite of
the presence of additional color symmetries.  In particular, we
observe the familiar depletion near $z = 0$ for $N_c =3$.

The spectrum in Fig.~2 is flatter than that for $N_c = 3$. This is a
consequence of the greater motion of the individual eigenvalues in the
model with $N_c = 2$. This difference is familiar in chiral ensembles,
where it is interpreted as a consequence of the fact that the spectrum
of a real matrix ($\chi$GOE) is less rigid than that of a complex
matrix ($\chi$GUE). It is interesting to see that the level motion in
our model with $N_c = 2$ agrees with that expected for $\chi$GOE even
though the matrix elements of $W$ are no longer real.  Again, the
correlations induced by chiral symmetry are maintained in spite of the
additional constraints introduced by the color structure of $W$.

Of course, the preservation of the $N_c = 2$ microscopic spectral 
density in spite of the complexity of $W$ should not come as a surprise.  
The interactions in Eq.~(\ref{Zcolor}) are, in fact, pseudoreal when
$N_c = 2$~\cite{VanJac01}. Thus, we could have used an appropriate basis 
of states such that all matrix elements have vanishing imaginary parts.  
The model would then have displayed a greater similarity to a $\chi$RMT 
model with $\beta = 1$, and it would thus be natural to find a spectrum 
which reproduces that of $\chi$GOE.

\section{Other spectral properties}
\label{s:others}

We have also studied short range eigenvalue correlations by examining
the distribution of level spacings, $p(s)$.  Here, $s$ is the nearest
neighbor spacing measured in units of the local average spacing.
(Hence, the average value of $s$ is $1$ by definition.) Note that
$p(s)$ is a bulk observable which is sensitive to chiral symmetry only
for small $s$.  Figure 3 shows the relative difference between the
results obtained for the model of I with $N_c = 3$ and those of the
$\chi$GUE. The data were obtained from a set of nine independent
series of diagonalizations of 10000 matrices of size $N \times N = 120
\times 120$. To avoid side effects due to the finite range of the
random matrix support, we only kept a central portion of the spectrum
(30th eigenvalue to 90th). The level spacing distribution for a given
model is taken as a bin-by-bin average over the nine series of runs,
with error bars corresponding to the variance in each bin. The plot
shows $\delta p(s) = [p_{\chi{\rm GUE}}(s) - p_{N_c =
3}(s)]/p_{\chi{\rm GUE}}(s)$, where $p_{\chi{\rm GUE}}(s)$ is the
level spacing distribution for $\chi {\rm GUE}$ and $p_{N_c = 3}(s)$
is that for model I with $N_c = 3$.  [Note that $p_{\chi{\rm GUE}}(s)
= p_{{\rm GUE}}(s)$.]  The level spacing difference is everywhere
consistent with zero.  Similar results were obtained for a comparison
with $\chi$GOE.  Thus, the level spacing distribution reveals no
statistically significant differences between the models of
Refs.~\cite{VanJac99} and their corresponding chiral ensembles.

It is also useful to construct the variance-covariance matrix $D$,
defined as
\begin{equation}
D_{ij} = \langle (\lambda_i - {\bar \lambda_i})
(\lambda_j - {\bar \lambda_j}) \rangle,
\end{equation}
where ${\bar \lambda_i}$ is the mean value of eigenvalue $i$.
Numerical studies of the models of Refs.~\cite{VanJac99} indicate that
all eigenvalues of $D$ are greater than zero.  Thus, the eigenvalues
of the Dirac operator do not satisfy any linear constraints.  The
eigenvalues of $D$ are related to the statistically independent
fluctuations of the eigenvalues of the Dirac operator about their mean
values.  (This relation would be exact if the joint probability
distribution for the eigenvalues of the Dirac operator were strictly
Gaussian.)  Numerical studies reveal good agreement with the spectrum
of normal modes known analytically for the $\chi$GOE and
$\chi$GUE~\cite{OswSpl01}.  This provides an additional indication
that the inclusion of color symmetry has not altered the statistical
properties of the models of Refs.~\cite{VanJac99} from those of the
corresponding chiral ensembles.

\section{Conclusions} 
\label{s:concl}

We have investigated a number of properties of the eigenvalue spectrum
of random matrix models which include both chiral and color
symmetries.  We find no deviation from the analytic results of
$\chi$RMT for either the microscopic spectral density or the level
spacing distribution.  Given the relatively elaborate block structure
of these models, such studies are most easily performed numerically.
While more complicated spectral correlators have not been
investigated, we find no grounds to doubt that they will also reflect
the underlying chiral structure of the problem.  (It should be
emphasized that this extended block structure does not lead to
complicated forms for the partition function and that studies of the
macroscopic properties of these models are straightforward for all $T$
and $\mu$.)  The spin and color block structures of the interactions
do not seem to upset spectral features associated with chiral
symmetry.  Rather, these additional correlations appear to act in a
channel which is ``orthogonal'' to the chiral channel.  We do not
regard this result as surprising. One may think of the models of
Refs.~\cite{VanJac99} as schematic lattice calculations which
implement single-gluon exchange between discrete quark fields. This
``lattice calculation'' does not possess any other length scale than
that introduced by $\Sigma$ and is, hence, free of a Thouless
energy. Because chiral symmetry is respected by the interactions, the
results of $\chi$RMT should be anticipated at all small energies.
This is indeed what our numerical analysis reveals.

The inclusion of color does, of course, introduce additional
correlations in the spectrum of the Dirac operator.  It would be of
interest to construct new spectral measures which could probe these
correlations.  In particular, it would be valuable to find a color
analogue of the Banks-Casher relation to facilitate the determination
of the strength of the diquark condensate.

\section*{Acknowledgments}

We thank J. J. M. Verbaarschot for provocative questions which
initiated this work and D. Toublan and K. Splittorff for useful
discussions.

\newpage

\begin{center}
\begin{figure}
\includegraphics{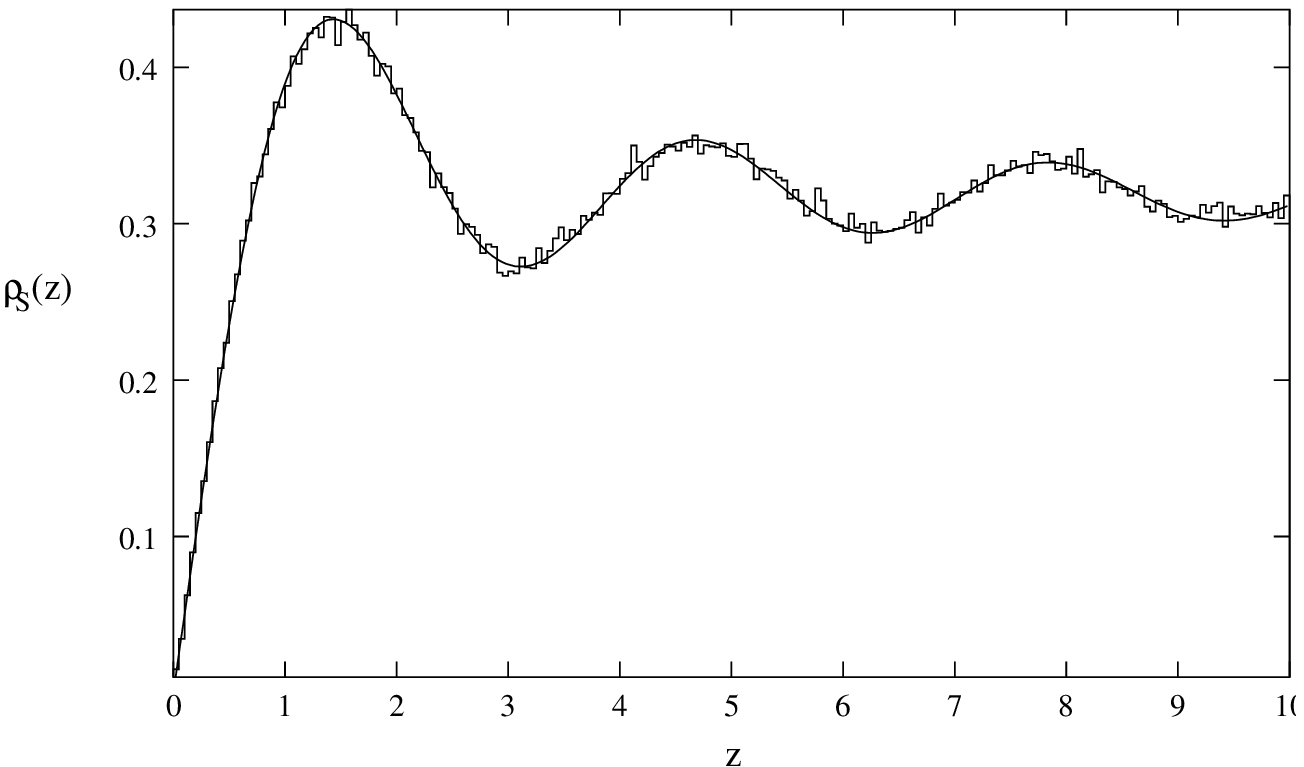}
\caption{Histogram: microscopic spectral density of the Dirac operator for a 
random matrix model with a color subblock and $N_c = 3$. Solid line: 
spectral density predicted for $\chi$GUE ($N_f \to 0$).}
\end{figure}
\end{center}

\newpage

\begin{center}
\begin{figure}
\includegraphics{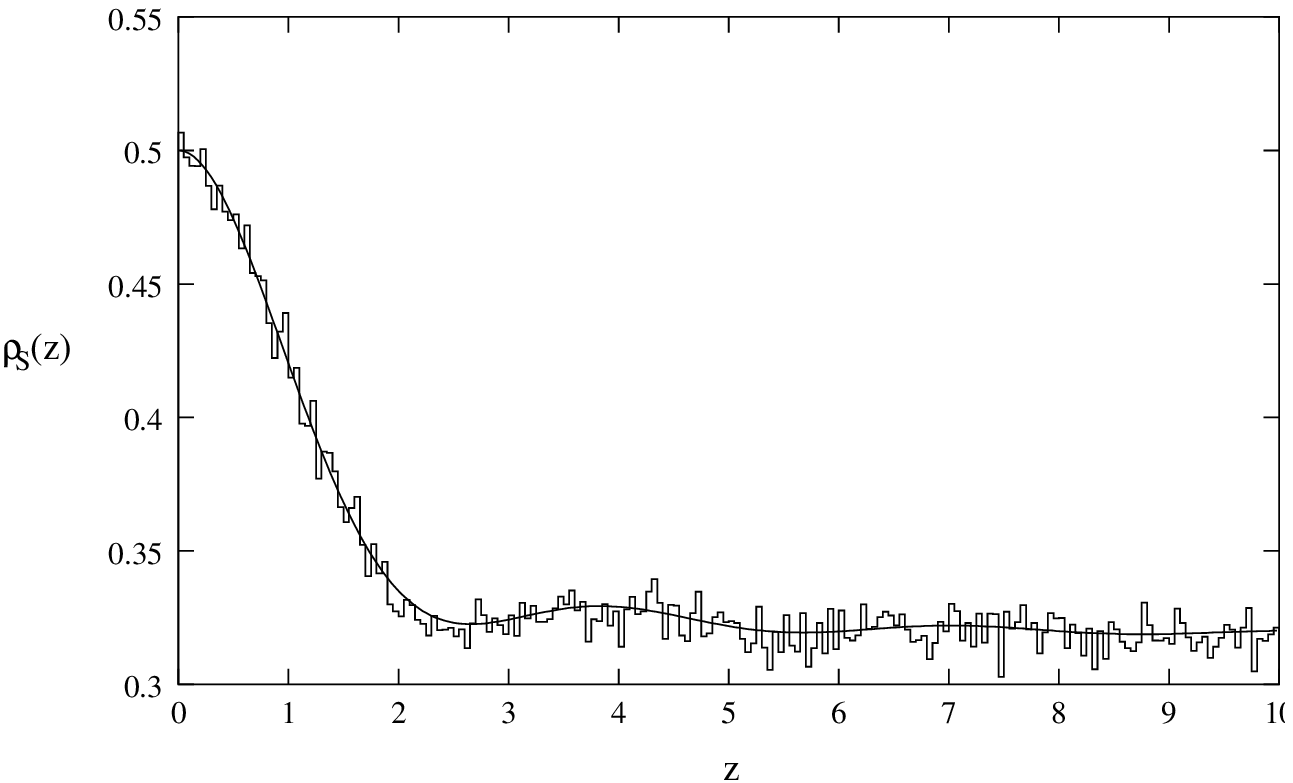}
\caption{Histogram: microscopic spectral density of the Dirac operator for a 
random matrix model with a color subblock and $N_c = 2$. Solid line: 
spectral density predicted for $\chi$GOE ($N_f \to 0$).}
\end{figure}
\end{center}

\newpage

\begin{center}
\begin{figure}
\includegraphics{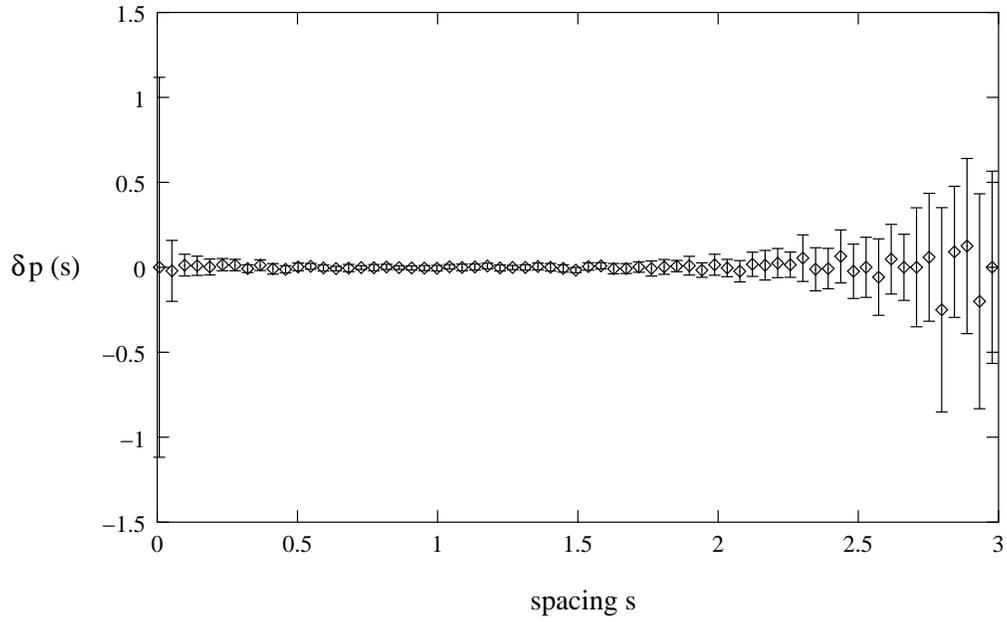}
\caption{Relative difference between the mean level spacing distribution 
for $\chi$GUE and that for a model with a color subblock and $N_c = 3$.
The errorbars are estimated from a series of nine independent 
runs of diagonalizations.}
\end{figure}
\end{center}

\end{document}